\documentclass[longbibliography,aps,prb,twocolumn,superscriptaddress]{revtex4-2}

\usepackage[colorlinks=true,linkcolor=blue,citecolor=blue,urlcolor=blue]{hyperref}
\usepackage{amsmath}
\usepackage{graphicx}
\usepackage{bm}
\usepackage{comment}
\usepackage{braket}

\begin{document}
\title{Surface-bulk hybridization enhances the Berry curvature dipole in GaAs(110)} 
\author{Gastón Blatter}
\affiliation{Institute for Theoretical Solid State Physics, IFW Dresden, Helmholtzstrasse 20, 01069 Dresden, Germany}
\author{Jorge I. Facio}
\affiliation{ Centro Atómico Bariloche, Instituto de Nanociencia y Nanotecnología (CNEA-CONICET) and Instituto Balseiro, Bariloche, Argentina}

\author{Jeroen van den Brink}
\affiliation{Institute for Theoretical Solid State Physics, IFW Dresden, Helmholtzstrasse 20, 01069 Dresden, Germany}
\affiliation{Würzburg-Dresden Cluster of Excellence ctd.qmat, Germany}

\date{\today} 

\begin{abstract}
Symmetry breaking at crystalline boundaries can enable electronic responses that are forbidden in the bulk. Here, we investigate the Berry curvature dipole (BCD) at GaAs(110) surfaces using first-principles calculations. While the BCD vanishes in bulk GaAs due to symmetry constraints, the reduced surface symmetry permits a finite dipole. We show that surface relaxation promotes hybridization between bulk and surface states, which strongly influences the surface BCD. Specifically, in unrelaxed slabs, the BCD arises predominantly from interband coupling among exponentially localized surface states lying within the projected bulk band gap. Surface relaxation shifts these states into the energy range of the bulk continuum, transforming them into surface resonances and promoting hybridization with bulk-like states. This hybridization generates pronounced Berry-curvature hot spots, resulting in an overall enhancement of the BCD. Our results highlight the importance of surface relaxation for accurately describing surface--bulk hybridization, which in turn plays a key role in Berry-curvature-driven responses at crystalline boundaries.
\end{abstract}


\maketitle

\section{ Introduction}

In crystalline materials, several electronic properties  are linked to the geometric and topological properties of the electronic band structure~\cite{Provost1980,Niu1999,Sinova2007,Nagaosa2010,Xiao2010,Neupert2013,Ding2022,Du2019,Nandy2019,Zhang2021,Du2021,Xie2021,Agarwal2022,Agarwal2025}. Starting from the quantum geometric tensor,  different transport and optical phenomena find a source in either the real part of this tensor, i.e., the quantum metric, or its imaginary part, known as the Berry curvature (BC). 

The BC can give rise to the so-called anomalous Hall effect, which produces a Hall current linear in the applied electric field, provided that time-reversal symmetry $\mathcal{T}$ is broken. 
In contrast, in $\mathcal{T}$-invariant noncentrosymmetric materials, a finite BC is allowed, but the linear Hall response vanishes. In this case, the leading contribution to the Hall response arises at second order in the electric field~\cite{Sodemann2015}. The so-called nonlinear Hall effect is directly linked to the first moment of the BC, known as the Berry curvature dipole (BCD). 

Since the BCD is strongly constrained by the symmetries of the underlying point group, low-symmetry materials are natural candidates for hosting nonlinear Hall effects in the bulk~\cite{Ma2019-qi,Kang2019-ek,PhysRevB.97.041101,PhysRevLett.121.246403,PhysRevLett.121.266601,PhysRevB.98.121109,https://doi.org/10.1002/qute.202100056}. Moreover, surfaces may naturally break symmetries, allowing for a finite BCD even when it is forbidden in the bulk~\cite{Wawrzik2021,Ovalle2022,PhysRevResearch.5.033007,wawrzik2023surface,PhysRevLett.134.066603}. Furthermore, while the origin of the BC is typically associated with internal degrees of freedom, such as spin, orbital, or site degrees of freedom, the presence of a surface introduces an additional source of BC through the dependence of the penetration depth of the Bloch wave function on the conserved crystalline momenta. For example, in models of Weyl semimetals, the delocalization of surface states conforming the Fermi arcs has been shown to lead to divergences of the BC, thereby enhancing the BCD as well~\cite{Wawrzik2021,PhysRevResearch.5.033007}.

Gallium arsenide (GaAs) is a prototypical III–V semiconductor with direct relevance for electronic devices~\cite{PhysRevLett.79.1917, Rössler_2010, SCHLAPFER2016114, PhysRevLett.117.256601, doi:10.1126/science.1216022, Külah_2021}, whose bulk phase crystallizes in space group 216 (zincblende structure)~\cite{Blakemore1982}. Although this structure lacks inversion symmetry, the BCD is forbidden by the combined action of its space-group symmetries~\cite{Gallego2019}. At the same time, GaAs and related III–V compounds are well suited for thin-film and surface studies, as they can be reliably cleaved along several crystallographic directions. Moreover, it can be grown and carefully controlled by Molecular Beam Epitaxy and allows the gating of heterostructures and quantum wells with 2D electron gasses (2DEG). In particular, the (110) surface is of special interest, since the symmetry is reduced to space group 31, resulting in a noncentrosymmetric structure with lower symmetry that allows for a finite BCD.

The GaAs (110) surface has been extensively investigated both experimentally and theoretically~\cite{Chang1984,Nienhaus1997,Duke1985,Lubinsky1976,Masud1982,Yi2011,Joannopoulos1974,Sabisch1995,Engels1998,Kulkova2005,Jiang2020,Garcia-Moliner1976,Kahn1978,Qian1988}, and its electronic structure is known to be strongly affected by surface relaxation. In particular, relaxation can induce substantial changes in the localization of surface states~\cite{Engels1998}. These effects are expected to have a significant impact on band-geometric properties such as the BCD and the Berry curvature~\cite{Wawrzik2021}.

In this work, we systematically investigate the electronic structure and Berry curvature dipole of GaAs(110) slabs of varying thickness, focusing on the role of surface states and structural relaxation. Using first-principles calculations, we analyze how symmetry and surface relaxation govern nonlinear transport responses in this prototypical semiconductor. Specifically, we show that the well-known reconstruction of the GaAs(110) surface strongly enhances surface--bulk hybridization, which in turn has a profound impact on the surface BCD. 

This article is organized as follows. In Section \ref{sec:Materials_methods}, we describe the methods used throughout the work. In Section~\ref{sec:band_structure}, we study the electronic band structure  of GaAs(110) with and without surface relaxation. In Section~\ref{sec:BCD_in_GaAs}, we present calculations of the the BCD in GaAs, focusing on the effects of the surface relaxation. Lastly, in Section~\ref{sec:conclusions}, we present our concluding remarks.

\section{Methods}
\label{sec:Materials_methods}

\subsection{Density-functional calculations}

\begin{figure}[h]
    \centering
    \includegraphics[width=\linewidth]{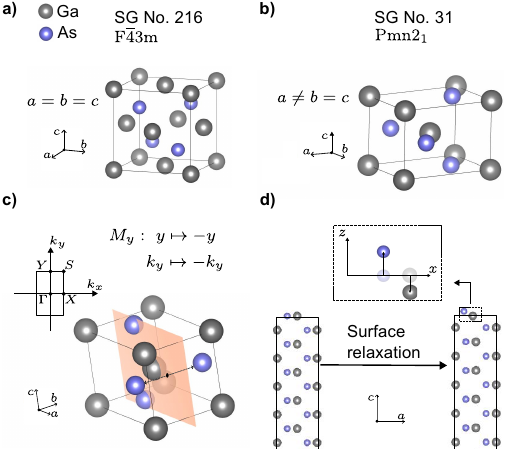}
    \caption{Structural and symmetry properties of bulk and (110)-oriented GaAs. (a) Crystal structure and space group of bulk GaAs. (b) Crystal structure and space group of the GaAs(110) surface. (c) Brillouin zone and mirror symmetry ($M_y$) preserved at the GaAs(110) surface. (d) Atomic displacement induced by surface relaxation.}
    \label{fig:fig1}
\end{figure}

To explore the physics of the (110) surface, we investigated the electronic band structure of GaAs slabs with surfaces perpendicular to the bulk (110) direction. The natural unit cell for constructing such slabs is no longer cubic but tetragonal, with lattice parameters satisfying $a\neq b = c$, where $c$ is perpendicular to the (110) surface~\cite{Aroyo2006}. The finite slabs can be constructed such that they belong to space group P$mn2_1$. The corresponding crystal structures are shown in Fig.~\ref{fig:fig1}(a) and (b). Although the symmetry is reduced relative to the bulk, the slab space group still contains a mirror plane,  illustrated in Fig.~\ref{fig:fig1}(c), which  constrains the BCD to lie perpendicular to it.

We constructed supercells with slab thicknesses $t=nc$, where $n=6, 8, 10,$ and $12$. A vacuum region was introduced along the $c$ direction to decouple periodic images of the slab. A vacuum thickness of $15$~\AA\ was found to be sufficient to suppress spurious interactions between adjacent slabs. The use of a larger vacuum leads to differences in the total energy of the order of $10^{-9}$~eV. 

We performed density functional theory (DFT) calculations using the FPLO package~\cite{Koepernik1999,Opahle1999} (version 22.02) within the generalized gradient approximation (GGA). For the self-consistent calculation of the charge density, we employed a $\mathbf{k}$-mesh of $12 \times 12 \times 1$ subdivisions, which was found to be sufficient to converge the total energy.

To assess the effects of surface relaxation on the electronic structure, structural optimizations were performed for the first surface layer. The atomic positions were relaxed by minimizing the forces within the scalar relativistic approximation with a tolerance of $1\times10^{-3}$ eV/\AA. Subsequent electronic structure calculations were performed with the fully-relativistic four-component scheme implemented in FPLO.
We have checked that relaxing deeper layers  only introduce moderate corrections to the surface atomic reconstruction.

Structurally, the relaxation results in the As atom being displaced outward from the slab, while the Ga atom is shifted toward the interior of the slab.  This is illustrated schematically in Fig.~\ref{fig:fig1}(d).  According to our calculations, the angle of rotation of the surface Ga-As bond is of approximately $28^o$, consistent with what is reported in the literature for this material ~\cite{Tong1978,duke1983,Smit1985}. 
The repository ~\cite{Blatter2026} provides the different structural models used throughout this work.

For each slab thickness, tight-binding models were constructed using symmetry-conserving projected Wannier functions based on the Ga and As $4s$ and $4p$ orbitals, as implemented in pyFPLO~\cite{Koepernik2023}. These models accurately reproduce the low-energy electronic structure obtained from DFT, with a calculated standard deviation of approximately $6$~meV in the energy window ranging from $-1$ to $1$~eV. These models were used to compute the BC and perform further postprocessing analyses of the band structure.

\subsection{Berry curvature dipole}

The electronic structure is described by the Wannier Hamiltonian
$H(\mathbf{k})$ obtained from the projective Wannier construction described above. Its eigenstates satisfy
\begin{equation}
H(\mathbf{k})\ket{n,\mathbf{k}}=
\varepsilon_n(\mathbf{k})\ket{n,\mathbf{k}},
\end{equation}
where $\ket{n,\mathbf{k}}$ and $\varepsilon_n(\mathbf{k})$ denote, respectively, the Bloch eigenstate and energy of band $n$ at crystalline momentum $\mathbf{k}$. The velocity operators are defined as
\[
v_a(\mathbf{k})=\frac{1}{\hbar}\frac{\partial H(\mathbf{k})}{\partial k_a},
\]
where, owing to the two-dimensional periodicity of the slab geometry, $a=x,y$.

The Berry curvature is a band-geometric quantity that characterizes the local geometry of the Bloch wave functions in momentum space. For the $n$th band, its $z$ component can be expressed as
\begin{equation}
\Omega_n^z(\mathbf{k}) =
-\mathrm{Im}\sum_{m\neq n}
\frac{
\braket{n,\mathbf{k}|v_x|m,\mathbf{k}}
\braket{m,\mathbf{k}|v_y|n,\mathbf{k}}
}{
[\varepsilon_n(\mathbf{k})-\varepsilon_m(\mathbf{k})]^2
},
\label{eq_bc}
\end{equation}
where the sum runs over all bands $m\neq n$ at the same crystalline momentum $\mathbf{k}$.

The Berry curvature dipole (BCD) is a pseudotensor that characterizes the first moment of the BC in momentum space~\cite{Sodemann2015}. It is defined as
\begin{equation}
D_{ab}=
\sum_n\int\frac{d^d k}{(2\pi)^d}f_0(\varepsilon_n)\,\partial_a\Omega_n^b(\mathbf{k}),
\label{eq}
\end{equation}
where $d$ denotes the dimensionality of the system, $f_0$ is the Fermi--Dirac distribution, and $\partial_a\equiv\partial/\partial k_a$. Introducing the shorthand
\begin{equation}
\int_{\mathbf{k}}\equiv\sum_n\int\frac{d^d k}{(2\pi)^d},
\end{equation}
the BCD can be written compactly as
\begin{equation}
D_{ab} =\int_{\mathbf{k}} f_0\,\partial_a\,\Omega_b.
\end{equation}
Integrating by parts yields
\begin{equation}
D_{ab}=-\int_{\mathbf{k}}\big(\partial_a f_0\big)\,\Omega_b,
\label{eq:BCD}
\end{equation}
which highlights the role of states near the Fermi surface. Indeed, $\partial_a f_0=(\partial f_0/\partial\varepsilon_n)\partial_a\varepsilon_n$, and in the zero-temperature limit $-\partial f_0/\partial\varepsilon_n$ approaches a Dirac delta function centered at the Fermi energy. 

To analyze the spatial distribution of the BCD across a slab, we implemented a Wannier projection scheme for the BCD. At each $\mathbf{k}$ point, the Berry curvature of the Hamiltonian eigenstates was evaluated together with the velocity matrix elements obtained from derivatives of the Wannier Hamiltonian. The contribution of each Wannier function $i$ to the BCD was then obtained by projecting the band-resolved BCD integrand onto the Wannier basis:

\begin{equation}
D_{ab}^{i} =
\int_{\mathbf{k}}
\sum_{n,j}
\Omega_{b,n}(\mathbf{k})
\frac{\partial f_0}{\partial \varepsilon_n}
 C^{*}_{i,n}(\mathbf{k})
 v_{a,ij}(\mathbf{k})
C_{j,n}(\mathbf{k}).
\end{equation}
Here, $C_{i,n}(\mathbf{k})$ are the eigenstate coefficients in the Wannier basis, $v_{a,ij}(\mathbf{k})$ denotes the velocity matrix element, and $i$ labels the Wannier function onto which the BCD contribution is projected. This decomposition provides an orbital- and site-resolved representation of the BCD, enabling the identification of the microscopic regions responsible for the nonlinear Hall response. By assigning each Wannier orbital to its corresponding atomic layer, we further obtain a layer-resolved decomposition of the BCD.

To identify the dominant interband processes contributing to the BCD, we further decompose the BC of each band into contributions arising from all other bands. From Eq.~\ref{eq_bc}, we define
\begin{equation}
\Omega_n^z(\mathbf{k})=\sum_{m\neq n}\Omega_{nm}^z(\mathbf{k}),
\label{eq}
\end{equation}
where
\begin{equation}
\Omega_{nm}^z(\mathbf{k})=-\mathrm{Im}\frac{\braket{n,\mathbf{k}|v_x|m,\mathbf{k}}\braket{m,\mathbf{k}|v_y|n,\mathbf{k}}}{[\varepsilon_n(\mathbf{k})-\varepsilon_m(\mathbf{k})]^2}
\end{equation}
denotes the contribution to the BC of band $n$ arising from band $m$. This decomposition allows us to express the contribution of band $n$ to the BCD as
\begin{equation}
D_{ab}^n=-\sum_{m\neq n}\int_{\mathbf{k}}\partial_a f_0\,\Omega_{nm}^b=
\sum_{m\neq n}D_{ab}^{n\leftarrow m}.
\label{eq:m_to_n}
\end{equation}
Here, $D_{ab}^{n\leftarrow m}$ denotes the contribution to the BCD of band $n$ arising from the interband contribution of band $m$.

Finally, for the numerical evaluation of the BCD, we used a 
$2000\times2000$ $\mathbf{k}$-point mesh. The integration was restricted to 
the first quadrant of the Brillouin zone, exploiting crystal and time-reversal 
symmetries to reconstruct the contributions from the remaining regions. 

\section{G\lowercase{A}A\lowercase{s}(110) Band Structure}
\label{sec:band_structure}
\begin{figure}
    \centering
    \includegraphics[width=\linewidth]{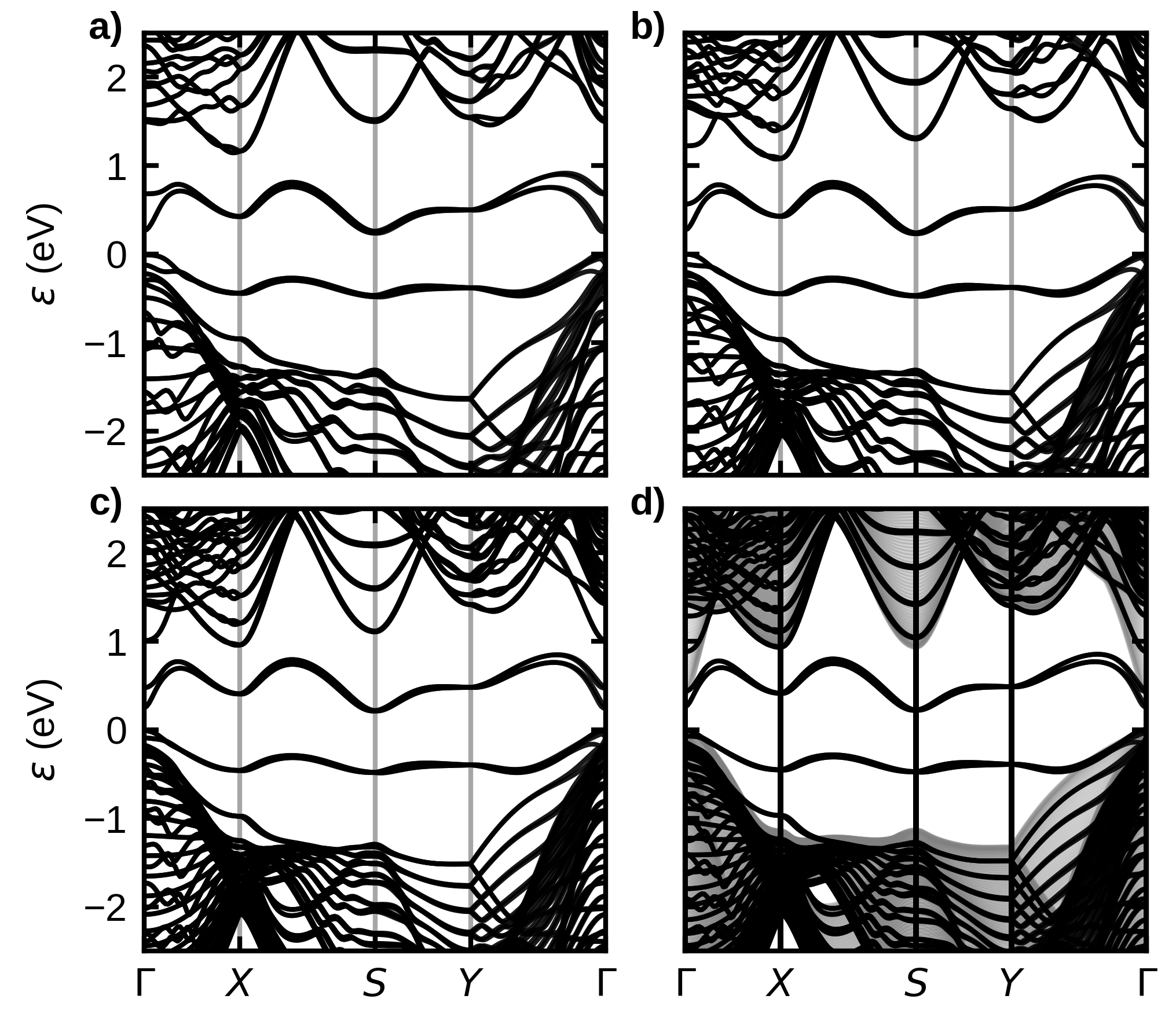}
    \caption{Electronic band structure of GaAs(110) slabs. (a–d) Electronic band structures for slab thicknesses \(n=6\), \(8\), \(10\), and \(12\), respectively, where \(n\) denotes the number of bulk unit cells forming the slab. In (d), the grey continuum spectrum corresponds to the bulk projection.}
    \label{fig:fig2}
\end{figure}

The electronic band structure of GaAs has been extensively studied, both theoretically and experimentally. GaAs is a semiconductor with an experimental direct band gap at $\Gamma$ of approximately $1.4$~eV~\cite{Blakemore1982}. The (110) surface is known to host electronic states within the bulk band gap, which is a characteristic feature of its surface electronic structure. To investigate these states, we computed the electronic band structure of GaAs slabs of varying thickness. As shown in Fig.~\ref{fig:fig2}, a set of states emerges above and below the Fermi level for all thicknesses considered. Importantly, the number of these states does not scale extensively with the number of layers $n$: only four of such states are observed for all values of $n$. This thickness-independent number of states suggests that they are associated with the surfaces of the slab.

For $n=12$, Fig.~\ref{fig:fig2} also shows the projected bulk band structure, obtained by varying the crystalline momentum along the direction perpendicular to the surface (110). In the limit $n\to\infty$, it is expected that a portion of the spectral weight of the slab converges to the continuum formed by the projected bulk states. Already for $n=12$, most of the slab band structure lies within the projected bulk bands, whereas the emergent states identified above remain within the bulk band gap.

As a quantitative measure of the spatial localization of a state, we introduce the surface polarization parameter
\begin{equation}
\eta(\mathbf{k}) = w_s(\mathbf{k}) - w_b(\mathbf{k}) = 2w_s(\mathbf{k}) - 1,
\label{eq}
\end{equation}
where $w_s(\mathbf{k})$ is the weight of a Bloch state on sites located near the surfaces of the slab, and $w_b(\mathbf{k})$ is the corresponding weight on intermediate, bulk-like sites. Owing to the normalization of the eigenstates, $w_s(\mathbf{k}) + w_b(\mathbf{k}) = 1$ for all $\mathbf{k}$ points in the Brillouin zone. With this definition, states fully localized at the surfaces have $\eta(\mathbf{k}) = 1$, whereas states fully localized in the bulk have $\eta(\mathbf{k}) = -1$. 
The surface region is defined as the first three atomic layers at both the top and bottom surfaces of the slab. 

\begin{figure}
    \centering
    \includegraphics[width=\linewidth]{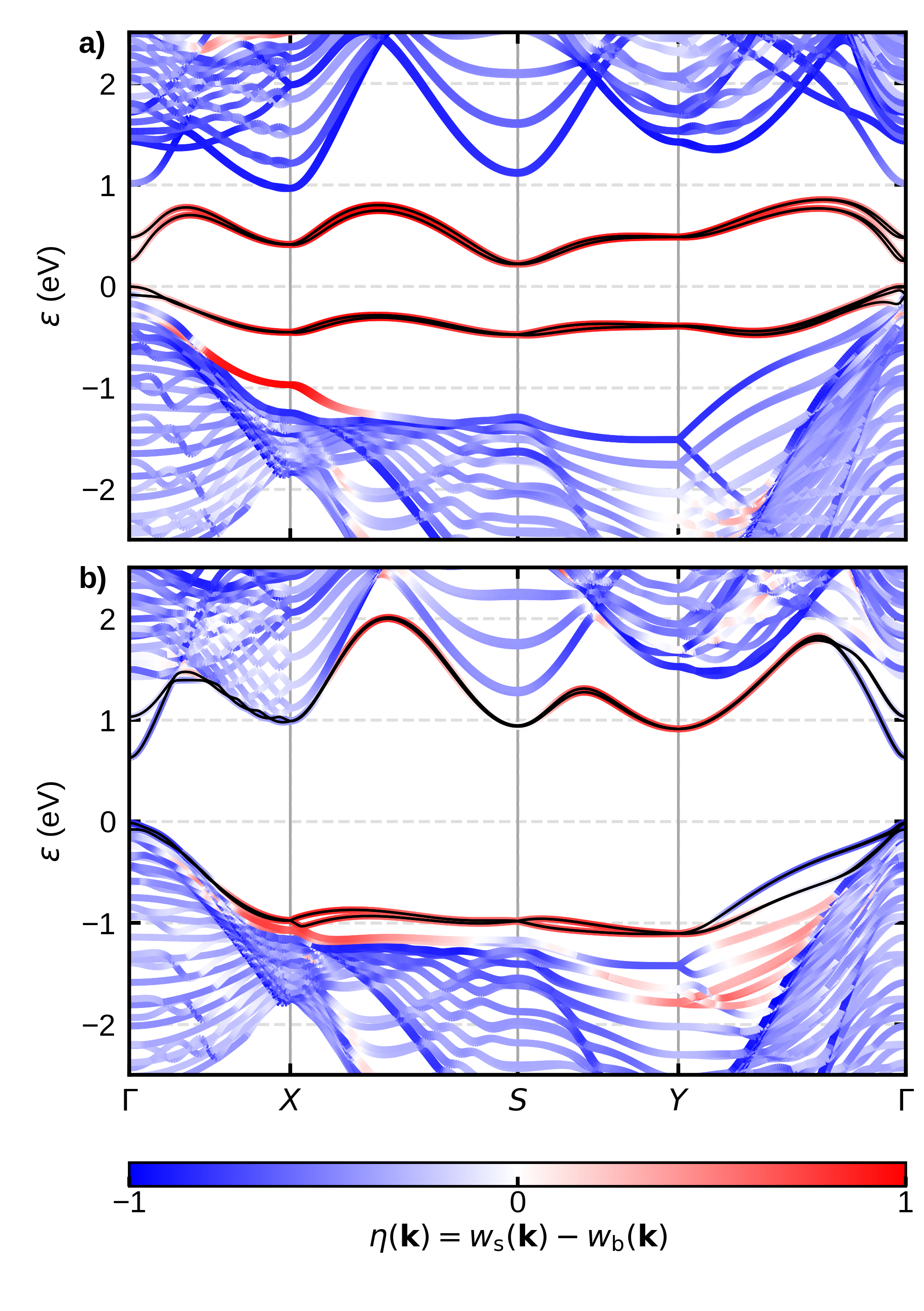}
    \caption{Band structure along the high-symmetry path for the $n=10$ slab. The colorbar indicates the surface polarization for (a) the slab without relaxation and (b) the slab with relaxed surface. The solid black lines highlight the states labeled $g_{ci}$ and $g_{vi}$ (see main text), which in (a) lie within the bulk band gap.}
    \label{fig:fig3}
\end{figure}

Fig.~\ref{fig:fig3}(a) presents the band structure of the $n=10$ slab, colored according to $\eta(\mathbf{k})$. The states lying within the band gap are strongly localized at the surfaces, with those in the immediate vicinity of the $\Gamma$ point, near the top of the bulk valence band, exhibiting a somewhat larger spread into the bulk. Throughout this article, we use the notation $g_{ci}$ ($i=1,2,3,4$) to denote the unoccupied states within the bulk band gap, and $g_{vi}$ to denote the corresponding occupied states. The remaining conduction and valence bands are indexed by $c_i$ (counted from the bottom of the conduction band) and $v_i$ (counted from the top of the valence band), respectively. Hereafter, bands $g_{ci}$ and $g_{vi}$ are indicated by solid black lines to distinguish them from the other bands.

We next consider the effects of surface relaxation, implemented as described in Section~\ref{sec:Materials_methods}.
The structural changes lead to significant modifications of the electronic band structure. In particular, the states that previously lay within the bulk band gap shift toward the energy range of the bulk states, as shown in Fig.~\ref{fig:fig3}(b).
Due to their shift in energy, at selected $\mathbf{k}$ points, these states strongly hybridize with the bulk continuum and therefore extend into the bulk. In the rest of this paper, we explain how this change affects BC-based properties.

\section{Berry Curvature Dipole in G\lowercase{A}A\lowercase{s}}
\label{sec:BCD_in_GaAs}

\subsection{Bulk and Surface GaAs BCD}
The BCD vanishes in bulk GaAs as a consequence of symmetry constraints. To determine which components of the BCD are allowed by symmetry, it is useful to recall that $D^{ab}$ is a second-rank pseudotensor: the index $a$, associated with the momentum derivative $\partial_a$, transforms as a polar vector, whereas the index $b$, associated with the Berry curvature $\Omega_b$, transforms as an axial vector. Consequently, the BCD transforms according to the direct product
$\Gamma_V\otimes\Gamma_A$, where $\Gamma_V$ and $\Gamma_A$ denote the vector and axial-vector representations of the crystal point group. A finite BCD is possible only if this product contains the trivial irreducible representation $A_1$, corresponding to a tensor component that remains invariant under all crystal symmetry operations.

For bulk GaAs, the point group is $\overline{4}3m$. The vector representation corresponds to the three-dimensional irreducible representation (irrep) $T_2$, whereas the axial-vector representation corresponds to $T_1$. Their direct product decomposes as
\begin{equation*}
\Gamma_V \otimes \Gamma_A
=
T_2 \otimes T_1
=
A_2 \oplus E \oplus T_1 \oplus T_2.
\end{equation*}
Since this decomposition does not contain the trivial irrep $A_1$, no component of the BCD remains invariant under all symmetry operations, and the BCD is therefore forbidden in bulk GaAs.

The situation changes at the (110) surface, where the symmetry is reduced to the point group $2mm$. In this case, the vector and axial-vector representations decompose as
\begin{equation*}
\Gamma_V = A_1(x) \oplus B_1(y) \oplus B_2(z),
\end{equation*}
and
\begin{equation*}
\Gamma_A = A_2(J_x) \oplus B_2(J_y) \oplus B_1(J_z),
\end{equation*}
respectively. Here, $(x,y,z)$ denote the components of a polar vector, whereas $(J_x,J_y,J_z)$ denote those of an axial vector. Since all irreducible representations of $2mm$ are one-dimensional, the product of a polar and an axial component transforms as the trivial representation only when both belong to the same irrep. Inspection of the above decompositions immediately shows that this occurs only for the pairs $(y,J_z)$ and $(z,J_y)$, giving
\begin{equation*}
A_1(y,J_z)\oplus A_1(z,J_y)
\subset
\Gamma_V\otimes\Gamma_A.
\end{equation*}
Accordingly, only the tensor components $D^{yz}$ and $D^{zy}$ are allowed by symmetry. For a two-dimensional slab, however, $k_z$ is not a good quantum number, so only $D^{yz}$ is physically relevant. Under these conditions, the BCD can be represented as an in-plane pseudovector with only  $D_y$ component allowed by symmetry.

Starting from the Wannier models of the slabs, we computed the $D_y$ component of the BCD for both unrelaxed and relaxed surface structures. At this point, it is useful to discuss the quantitative impact of the band gap on our calculations. While the experimentally reported band gap of GaAs is approximately $1.4$~eV~\cite{Blakemore1982}, DFT typically underestimates its magnitude~\cite{Perdew1983,Ma2016}, yielding significantly smaller values~\cite{Anua2012,Ahmed2009,Wang2002,Rushton2001,Kalvoda1997,Causa1991}. In principle, this discrepancy may affect the Berry curvature, which depends sensitively on the energy separation between bands, and consequently the magnitude of the BCD. However, the smallest interband gaps relevant to the features discussed below occur predominantly between conduction bands. We therefore expect the underestimated fundamental gap between the valence and conduction bands to play a less significant role in the qualitative behavior of the BCD considered here.

In addition, several other factors may affect the magnitude of the experimentally measured BCD and the associated nonlinear transport response, including disorder and the crystal domain structure. These factors make quantitatively accurate predictions of transport coefficients challenging. Since our analysis primarily focuses on the symmetry and microscopic origin of the BCD, we therefore present the results obtained directly from DFT and focus on a qualitative comparison of the changes induced by surface relaxation.

\subsection{Unrelaxed Slabs}

\begin{figure}
    \centering
    \includegraphics[width=\linewidth]{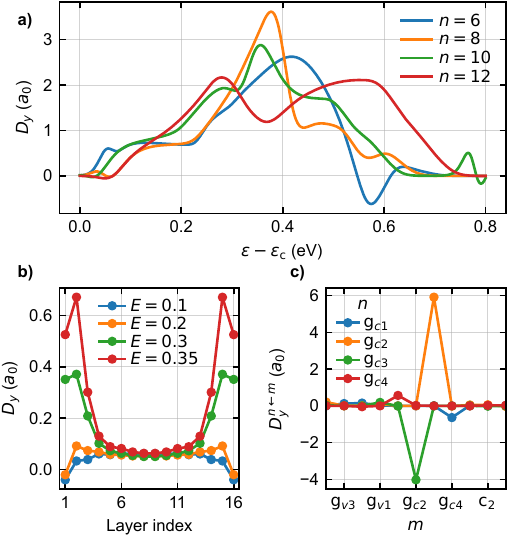}
    \caption{Berry curvature dipole (BCD) for the unrelaxed slabs. (a) Energy dependence of the BCD with respect to the bottom of the conducting band. (b) Layer dependence of the BCD for different energies with respect to $\varepsilon_c$. The unit for the energies is eV. (c) $D_y^{n\leftarrow m}$ at $\varepsilon-\varepsilon_c=0.35$ eV. We include only the $n$ that have a finite isoenergy surface at this energy, since those are the ones contributing to the BCD.}
    \label{fig:fig4}
\end{figure}
For unrelaxed slabs, we calculate the BCD within the energy window spanned by the conduction-band surface states, from $0$ to $0.8$~eV, as shown in Fig.~\ref{fig:fig4}(a) in units of Bohr radius $a_0$. To analyze the real-space distribution of the BCD, we resolve its contributions into individual atomic layers at fixed energy following the procedure described in Sec.~\ref{sec:Materials_methods}. The layer-resolved contributions for the $n=8$ slab are shown in Fig.~\ref{fig:fig4}(b). At energies close to the bottom of the conduction band ($\varepsilon-\varepsilon_c\lesssim0.2$~eV), the BCD is predominantly concentrated in the intermediate, bulk-like layers. As the energy increases, a surface-dominated contribution emerges, primarily localized within the outermost three layers. The intermediate layers retain a finite contribution, which we attribute to finite-size effects. Indeed, in the bulk limit, $n\to\infty$, this contribution is expected to vanish by symmetry as the full $\overline{4}3m$ point-group symmetry is recovered.

To understand the origin of the BC associated with the dominant surface contribution, we analyze the interband contributions to the BCD (Eq.~\ref{eq}), $D_y^{n\leftarrow m}$, shown in Fig.~\ref{fig:fig4}(c) for the $n=8$ slab at $\varepsilon-\varepsilon_c\sim0.35$~eV, where the BCD is predominantly localized at the surface. We find that the dominant contributions originate from interband matrix elements connecting surface-localized states, namely those lying within the bulk band gap.  This analysis shows that the BCD is generated primarily by interband coupling between surface states, with negligible contributions from bulk-like bands.

Returning to the discussion of the underestimated band gap in DFT, we note that the contribution of a band $m$ to the Berry curvature of band $n$ scales as $1/[\varepsilon_n(\mathbf{k})-\varepsilon_m(\mathbf{k})]^2$. The dominant contributions identified here arise from small energy separations among surface states rather than from transitions across the fundamental valence--conduction band gap. We therefore expect the DFT underestimation of the latter to have a comparatively minor impact on the qualitative behavior of the BCD discussed here.

\subsection{Relaxed Slabs}

\begin{figure}
    \centering
    \includegraphics[width=\linewidth]{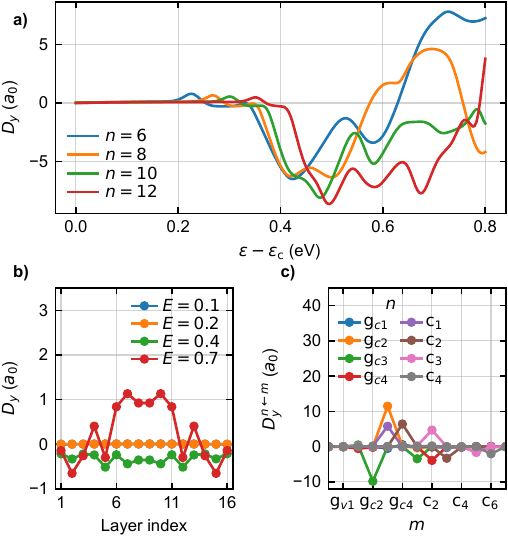}
    \caption{Berry curvature dipole (BCD) for the relaxed slabs. (a) Energy dependence of the BCD with respect to the bottom of the conducting band. (b) Layer dependence of the BCD for different energies with respect to $\varepsilon_c$. The unit for the energies is eV. (c) $D_y^{n\leftarrow m}$ at $\varepsilon-\varepsilon_c=0.7$ eV. We include only the $n$ that have a finite isoenergy surface at this energy, since those are the ones contributing to the BCD.}
    \label{fig:fig5}
\end{figure}

The strong band-structure reconstruction induced by surface relaxation is directly reflected in the BCD. As shown in Fig.~\ref{fig:fig5}(a), the BCD is strongly suppressed at low energies, between $0$ and $0.3$~eV, reaching values of only $\sim 0.66 \ a_0$. In this energy range, the conduction states retain a predominantly bulk-like character, as illustrated in Fig.~\ref{fig:fig5}(b), resulting in $D_y(\varepsilon)$ profiles that are nearly independent of slab thickness $n$, apart from an approximately rigid energy shift. At higher energies, the BCD increases significantly, reaching absolute values of $5$--$8 \ a_0$, approximately twice those obtained for unrelaxed slabs. This enhancement coincides with the onset of strong hybridization between surface-resonant states and an increasing number of bulk-like bands. Consequently, the detailed shape of $D_y(\varepsilon)$ becomes increasingly dependent on $n$, since both the number and energy distribution of the hybridizing bulk-like bands vary with slab thickness. Nevertheless, the overall magnitude of the dipole remains comparable for all values of $n$ considered. We therefore expect the BCD profile to converge in the limit of sufficiently thick slabs, although reaching this regime is currently limited by the computational cost of the calculations.

We have verified that relaxing additional atomic layers does not qualitatively modify these conclusions. For energies below $\varepsilon-\varepsilon_c=0.4$~eV, the BCD profiles closely resemble those shown in Fig.~\ref{fig:fig5}(a), apart from a small energy shift. At higher energies, however, the detailed profiles become increasingly dependent on the number of relaxed layers. As discussed above, we attribute this sensitivity to the large number of slab bands present in this energy window: even small modifications of the electronic structure can produce substantial changes in geometric quantities such as the BC and the BCD. This sensitivity is an inherent consequence of describing bulk continuum states within a finite slab geometry. Nevertheless, the overall magnitude of the BCD remains comparable for all relaxation depths considered, indicating that the main physical conclusions are robust against the number of relaxed layers.

The layer-resolved BCD shown in Fig.~\ref{fig:fig5}(b) also exhibits pronounced differences from the unrelaxed case. While in unrelaxed slabs the largest contributions originate from the surface layers, after relaxation the BCD is distributed throughout the entire slab. Consistently, the interband contributions to the BCD are also distributed among a broader set of bands than in the unrelaxed case, as shown in Fig.~\ref{fig:fig5}(c).

\begin{figure*}[t]
    \centering
    \includegraphics[width=\linewidth]{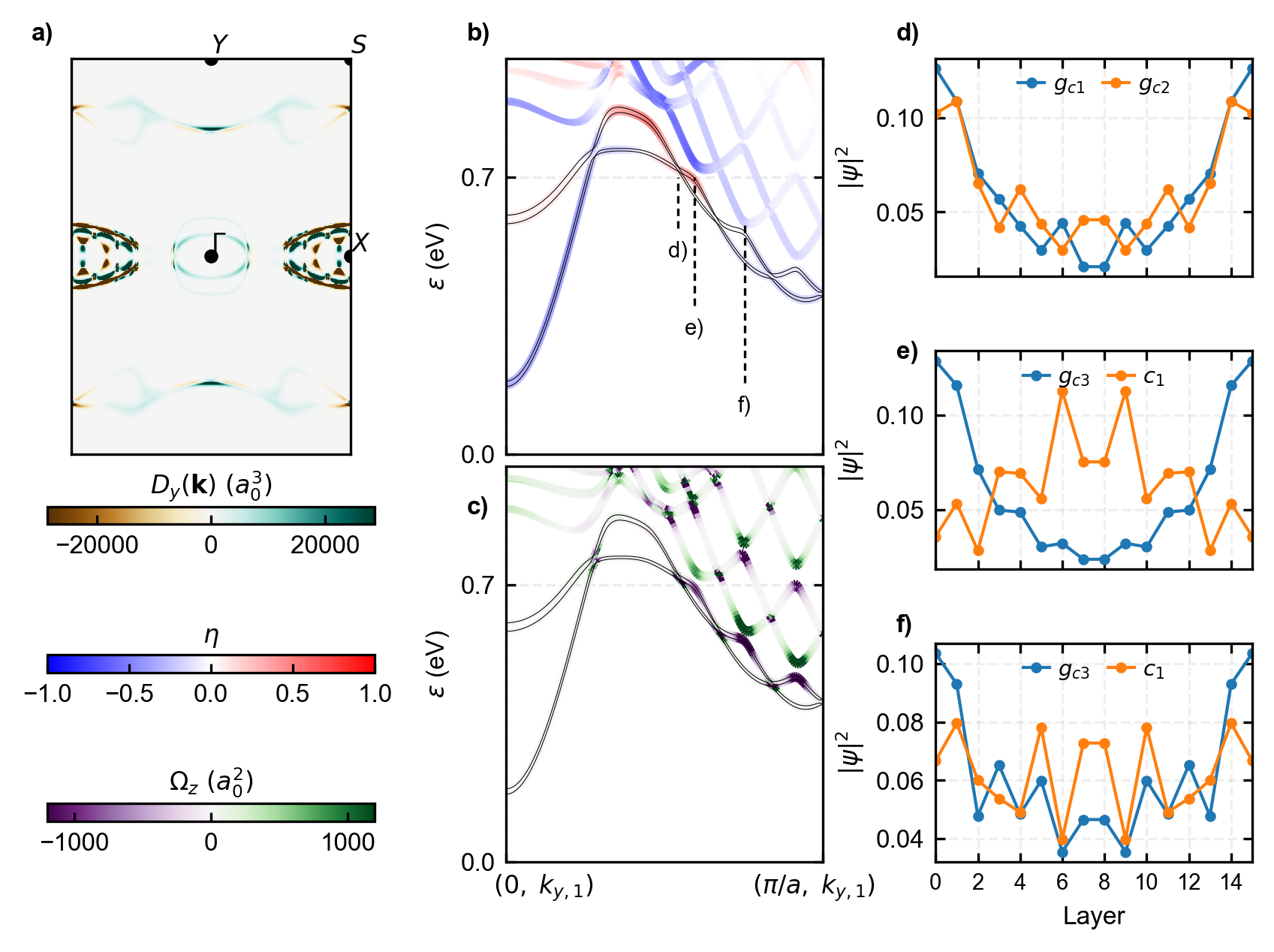}
    \caption{Momentum-space and layer-resolved characterization of the avoided crossings in the \(n=8\) slab.(a) Momentum-resolved distribution of the Berry-curvature dipole density $D_y(\mathbf{k})$ for the $n=8$ slab over the first Brillouin zone. The color scale is saturated at the 98th percentile of the data. (b) Band structure along the line from $(0,k_{y,1})$ to $(\pi/a,k_{y,1})$, with fixed $k_{y,1}=0.03 \ a_0^{-1}$. Color encodes the surface polarization $\eta$; the dashed vertical lines mark three representative avoided crossings. (c) Same band structure as in (b), with color indicating the Berry curvature $\Omega_z$. (d–f) Layer-resolved probability weight of the Bloch states forming, respectively, the left, middle, and right avoided crossings highlighted in (b). The layer profiles reveal the spatial localization of the states involved in different avoided crossings.}
    \label{fig:fig6}
\end{figure*}

\subsection{Hybridization of surface and bulk states}

To analyze the momentum-space origin of the BCD across the entire BZ, we study the $\mathbf{k}$-resolved density $D_y(\mathbf{k})$, which corresponds to the integrand in equation~\ref{eq:BCD}, for the $n=8$ slab. At $\varepsilon-\varepsilon_c=0.7$~eV, where $D_y$ reaches its largest value, Fig.~\ref{fig:fig6}(a) reveals strong contributions along paths parallel to the $\Gamma$--$X$ direction. The symmetry properties of $D_y(\mathbf{k})$ are consistent with those of the slab. First, time-reversal symmetry imposes $D_y(-\mathbf{k})=D_y(\mathbf{k})$. Second, the mirror symmetry $M_y$ requires $D_y(k_x,-k_y)=D_y(k_x,k_y)$. Combining these two constraints further yields $D_y(-k_x,k_y)=D_y(k_x,k_y)$.

The BCD hot spots identified in panel (a) coincide with localized enhancements of the Berry curvature associated with avoided crossings. To illustrate their microscopic origin, Fig.~\ref{fig:fig6}(b) shows the band structure along a representative path crossing these hot spots, $(0,k_{y,1})\rightarrow(\pi/a,k_{y,1})$, with fixed $k_{y,1}=0.03,a_0^{-1}$, with the bands colored according to their surface polarization $\eta(\mathbf{k})$. Figure~\ref{fig:fig6}(c) shows the same band structure colored by the Berry curvature $\Omega_z(\mathbf{k})$. The localized peaks in $\Omega_z$ occur at avoided crossings between states with different degrees of surface localization, three representative examples of which are marked by the dashed lines in panel (b).

The layer-resolved wave-function weights shown in Fig.~\ref{fig:fig6}(d)--(f) clarify the nature of these avoided crossings. Panel (d) illustrates an avoided crossing between two states that remain predominantly localized near the surfaces. Such surface--surface hybridizations are also present in the unrelaxed slabs and therefore do not constitute a qualitatively new mechanism introduced by relaxation. Surface relaxation, however, substantially changes the situation by shifting surface-derived bands into the energy range occupied by bulk states. As a result, the relaxed slabs exhibit a proliferation of avoided crossings involving states with markedly different spatial character. Panels (e) and (f) show representative examples: in each case, a state with pronounced surface weight hybridizes with a state whose wave function extends throughout the slab. These avoided crossings therefore correspond to surface--bulk hybridization.

This distinction is central to the enhancement of the BCD in the relaxed slabs. In the unrelaxed system, the relevant surface-derived states lie predominantly within the projected bulk gap, which restricts the number of bands with which they can hybridize. Surface relaxation shifts these states into the projected bulk continuum, where they become surface resonances and encounter a much denser set of bulk-like bands. The resulting proliferation of surface--bulk avoided crossings produces small interband energy separations and strong local enhancements of the Berry curvature. At the same time, a larger number of intermediate bands $m$ contribute appreciably to the BCD of a given band $n$ through Eq.~\eqref{eq:m_to_n}. The combined increase in the number of relevant interband couplings and the reduction of their characteristic energy denominators generates the pronounced BCD hot spots visible in Fig.~\ref{fig:fig6}(a) and accounts for the overall enhancement of the BCD upon surface relaxation.

The enhancement of the Berry curvature associated with the interplay between surface and bulk states is not unique to the present system. In Weyl semimetals, Fermi arcs have been shown to exhibit a divergent surface Berry curvature as they merge into the bulk states~\cite{Wawrzik2021}. Although GaAs does not host Fermi arcs and no such divergence occurs here, it is nevertheless noteworthy that an enhancement of the Berry curvature similarly emerges as surface-derived and bulk-like states approach each other in energy and hybridize.

\section{Conclusions}
\label{sec:conclusions}

We have investigated the microscopic origin of the Berry curvature dipole (BCD) in GaAs (110) slabs using density functional theory calculations combined with Wannier-function-based modeling. While the BCD vanishes in bulk GaAs due to symmetry constraints, the reduced symmetry at the surface allows for a finite dipole.

We have analyzed the differences between surfaces with and without structural relaxation, which is known to strongly affect the surface electronic structure of GaAs. In unrelaxed slabs, the BCD is mainly associated with exponentially localized surface states lying within the projected bulk band gap. Structural relaxation displaces the surface atoms, with As atoms shifting toward the vacuum and Ga atoms toward the interior of the slab. These structural changes shift the surface-derived states into the energy range of the bulk continuum and transform them into surface resonances. The resulting surface--bulk hybridization generates avoided crossings and redistributes the Berry curvature in momentum space, leading to a substantial enhancement of the BCD, which reaches values approximately twice as large as those obtained for unrelaxed slabs.

Our analysis further reveals that the dominant contributions arise from interband processes involving surface-derived states and are strongly enhanced near avoided crossings and regions of small interband energy separation. The enhancement of the BCD can thus be understood as the combined effect of the relaxation-induced delocalization of the surface states and their increased coupling to bulk-like states. More broadly, our results highlight the role of surface resonances as a source of enhanced Berry curvature and nonlinear transport responses.

GaAs (110) surfaces, and potentially other members of the III--V semiconductor family, therefore provide an attractive experimental platform for exploring Berry-curvature-driven responses associated with surface electronic states.
Since the strongest Berry-curvature features occur at energies beyond those readily accessible by conventional doping, our results also motivate the investigation of frequency-dependent nonlinear responses probing the same relaxation-induced surface--bulk hybridization~\cite{537j-v1lp,PhysRevLett.129.227401,nagaosa2017,sipe2000,yang2017,PhysRevX.10.041041}.

\section*{Acknowledgements} \label{sec:acknowledgements}
We thank U. Nitzsche for technical assistance, Xiao Zhang, Mikel Iraola, Diego García Ovalle, Jorge Cárdenas-Gamboa, Oleg Janson, Mengli Hu and Maia G. Vergniory for fruitful discussions.

\section*{Data availability}
The data supporting the findings of this study are openly available~\cite{Blatter2026}. The dataset includes the input files and converged charge densities used in the first-principles calculations, together with the information required to construct the Wannier models employed in this work.

\section*{Code availability}
The codes using to obtain the results presented in this article are available from the authors upon reasonable request.

\bibliography{references_v2} 

@article{Provost1980,
    author = "Provost, J. P. and Vallee, G.",
    title = "{Riemannian structure on manifolds of quantum states}",
    volume = "76",
    ISSN = "1432-0916",
    url = "http://dx.doi.org/10.1007/BF02193559",
    DOI = "10.1007/bf02193559",
    number = "3",
    journal = "Commun. Math. Phys.",
    publisher = "Springer Science and Business Media LLC",
    year = "1980",
    month = "",
    pages = "289–301"
}

@article{Neupert2013,
    author = "Neupert, Titus and Chamon, Claudio and Mudry, Christopher",
    title = "{Measuring the quantum geometry of Bloch bands with current noise}",
    volume = "87",
    ISSN = "1550-235X",
    url = "http://dx.doi.org/10.1103/PhysRevB.87.245103",
    DOI = "10.1103/physrevb.87.245103",
    number = "24",
    journal = "Phys. Rev. B",
    publisher = "American Physical Society (APS)",
    year = "2013",
    month = "",
    pages = "245103"
}

@article{Xiao2010,
    author = "Xiao, Di and Chang, Ming-Che and Niu, Qian",
    title = "{Berry phase effects on electronic properties}",
    volume = "82",
    ISSN = "1539-0756",
    url = "http://dx.doi.org/10.1103/RevModPhys.82.1959",
    DOI = "10.1103/revmodphys.82.1959",
    number = "3",
    journal = "Rev. Mod. Phys.",
    publisher = "American Physical Society (APS)",
    year = "2010",
    month = "",
    pages = "1959–2007"
}

@article{Nagaosa2010,
  title = {Anomalous Hall effect},
  author = {Nagaosa, Naoto and Sinova, Jairo and Onoda, Shigeki and MacDonald, A. H. and Ong, N. P.},
  journal = {Rev. Mod. Phys.},
  volume = {82},
  issue = {2},
  pages = {1539--1592},
  numpages = {0},
  year = {2010},
  month = {May},
  publisher = {American Physical Society},
  doi = {10.1103/RevModPhys.82.1539},
  url = {https://link.aps.org/doi/10.1103/RevModPhys.82.1539}
}

@article{Sodemann2015,
    author = "Sodemann, Inti and Fu, Liang",
    title = "{Quantum Nonlinear Hall Effect Induced by Berry Curvature Dipole in Time-Reversal Invariant Materials}",
    volume = "115",
    ISSN = "1079-7114",
    url = "http://dx.doi.org/10.1103/PhysRevLett.115.216806",
    DOI = "10.1103/physrevlett.115.216806",
    number = "21",
    journal = "Phys. Rev. Lett.",
    publisher = "American Physical Society (APS)",
    year = "2015",
    month = "November",
    pages = "216806"
}

@article{Ding2022,
    author = "Ding, Hai-Tao and Zhu, Yan-Qing and He, Peng and Liu, Yu-Guo and Wang, Jian-Te and Zhang, Dan-Wei and Zhu, Shi-Liang",
    title = "{Extracting non-Abelian quantum metric tensor and its related Chern numbers}",
    volume = "105",
    ISSN = "2469-9934",
    url = "http://dx.doi.org/10.1103/PhysRevA.105.012210",
    DOI = "10.1103/physreva.105.012210",
    number = "1",
    journal = "Phys. Rev. A",
    publisher = "American Physical Society (APS)",
    year = "2022",
    month = "January",
    pages = "012210"
}

@article{Ovalle2022,
  title = {Influence of the surface states on the nonlinear Hall effect in {Weyl} semimetals},
  author = {Ovalle, Diego Garc\'{\i}a and Pezo, Armando and Manchon, Aur\'elien},
  journal = {Phys. Rev. B},
  volume = {106},
  issue = {21},
  pages = {214435},
  numpages = {12},
  year = {2022},
  month = {Dec},
  publisher = {American Physical Society},
  doi = {10.1103/PhysRevB.106.214435},
  url = {https://link.aps.org/doi/10.1103/PhysRevB.106.214435}
}

@article{Gallego2019,
author = "Gallego, Samuel V. and Etxebarria, Jesus and Elcoro, Luis and Tasci, Emre S. and Perez-Mato, J. Manuel",
title = "{Automatic calculation of symmetry-adapted tensors in magnetic and non-magnetic materials: a new tool of the Bilbao Crystallographic Server}",
journal = "Acta Crystallogr. A",
year = "2019",
volume = "75",
number = "3",
pages = "438--447",
month = "May",
doi = {10.1107/S2053273319001748},
url = {https://doi.org/10.1107/S2053273319001748},
}

@article{Niu1999,
    author = "Sundaram, Ganesh and Niu, Qian",
    title = "{Wave-packet dynamics in slowly perturbed crystals: Gradient corrections and Berry-phase effects}",
    volume = "59",
    ISSN = "1095-3795",
    url = "http://dx.doi.org/10.1103/PhysRevB.59.14915",
    DOI = "10.1103/physrevb.59.14915",
    number = "23",
    journal = "Phys. Rev. B",
    publisher = "American Physical Society (APS)",
    year = "1999",
    month = "",
    pages = "14915–14925"
}

@article{Zhang2021,
    author = "Lai, Shen and Liu, Huiying and Zhang, Zhaowei and Zhao, Jianzhou and Feng, Xiaolong and Wang, Naizhou and Tang, Chaolong and Liu, Yuanda and Novoselov, K. S. and Yang, Shengyuan A. and Gao, Wei-bo",
    title = "{Third-order nonlinear Hall effect induced by the Berry-connection polarizability tensor}",
    volume = "16",
    ISSN = "1748-3395",
    url = "http://dx.doi.org/10.1038/s41565-021-00917-0",
    DOI = "10.1038/s41565-021-00917-0",
    number = "8",
    journal = "Nat. Nanotechnol.",
    publisher = "Springer Science and Business Media LLC",
    year = "2021",
    month = "",
    pages = "869–873"
}

@article{Du2021,
    author = "Du, Z. Z. and Wang, C. M. and Sun, Hai-Peng and Lu, Hai-Zhou and Xie, X. C.",
    title = "{Quantum theory of the nonlinear Hall effect}",
    volume = "12",
    ISSN = "2041-1723",
    url = "http://dx.doi.org/10.1038/s41467-021-25273-4",
    DOI = "10.1038/s41467-021-25273-4",
    number = "1",
    journal = "Nat. Commun.",
    publisher = "Springer Science and Business Media LLC",
    year = "2021",
    month = "August",
    pages = "5038"
}

@article{Xie2021,
    author = "Du, Z. Z. and Lu, Hai-Zhou and Xie, X. C.",
    title = "{Nonlinear Hall effects}",
    volume = "3",
    ISSN = "2522-5820",
    url = "http://dx.doi.org/10.1038/s42254-021-00359-6",
    DOI = "10.1038/s42254-021-00359-6",
    number = "11",
    journal = "Nat. Rev. Phys.",
    publisher = "Springer Science and Business Media LLC",
    year = "2021",
    month = "August",
    pages = "744–752"
}

@article{Agarwal2022,
    author = "Bhalla, Pankaj and Das, Kamal and Culcer, Dimitrie and Agarwal, Amit",
    title = "{Resonant Second-Harmonic Generation as a Probe of Quantum Geometry}",
    volume = "129",
    ISSN = "1079-7114",
    url = "http://dx.doi.org/10.1103/PhysRevLett.129.227401",
    DOI = "10.1103/physrevlett.129.227401",
    number = "22",
    journal = "Phys. Rev. Lett.",
    publisher = "American Physical Society (APS)",
    year = "2022",
    month = "November",
    pages = "227401"
}

@article{Agarwal2025,
    author = "Sarkar, Sanjay and Mandal, Debottam and Agarwal, Amit",
    title = "{Band geometry induced third-harmonic generation}",
    volume = "112",
    ISSN = "2469-9969",
    url = "http://dx.doi.org/10.1103/537j-v1lp",
    DOI = "10.1103/537j-v1lp",
    number = "24",
    journal = "Phys. Rev. B",
    publisher = "American Physical Society (APS)",
    year = "2025",
    month = "December",
    pages = "245425"
}

@article{Du2019,
    author = "Du, Z. Z. and Wang, C. M. and Li, Shuai and Lu, Hai-Zhou and Xie, X. C.",
    title = "{Disorder-induced nonlinear Hall effect with time-reversal symmetry}",
    volume = "10",
    ISSN = "2041-1723",
    url = "http://dx.doi.org/10.1038/s41467-019-10941-3",
    DOI = "10.1038/s41467-019-10941-3",
    number = "1",
    journal = "Nat. Commun.",
    publisher = "Springer Science and Business Media LLC",
    year = "2019",
    month = "",
    pages = "3047"
}

@article{Nandy2019,
    author = "Nandy, S. and Sodemann, Inti",
    title = "{Symmetry and quantum kinetics of the nonlinear Hall effect}",
    volume = "100",
    ISSN = "2469-9969",
    url = "http://dx.doi.org/10.1103/PhysRevB.100.195117",
    DOI = "10.1103/physrevb.100.195117",
    number = "19",
    journal = "Phys. Rev. B",
    publisher = "American Physical Society (APS)",
    year = "2019",
    month = "November",
    pages = "195117"
}

@article{Sinova2007,
    author = "Sinitsyn, N. A. and MacDonald, A. H. and Jungwirth, T. and Dugaev, V. K. and Sinova, Jairo",
    title = "{Anomalous Hall effect in a two-dimensional Dirac band: The link between the Kubo-Streda formula and the semiclassical Boltzmann equation approach}",
    volume = "75",
    ISSN = "1550-235X",
    url = "http://dx.doi.org/10.1103/PhysRevB.75.045315",
    DOI = "10.1103/physrevb.75.045315",
    number = "4",
    journal = "Phys. Rev. B",
    publisher = "American Physical Society (APS)",
    year = "2007",
    month = "January",
    pages = "045315"
}

@article{PhysRevResearch.5.033007,
  title = {{Berry curvature associated to Fermi arcs in continuum and lattice {Weyl} systems}},
  author = {Wawrzik, Dennis and van den Brink, Jeroen},
  journal = {Phys. Rev. Res.},
  volume = {5},
  issue = {3},
  pages = {033007},
  numpages = {9},
  year = {2023},
  month = {Jul},
  publisher = {American Physical Society},
  doi = {10.1103/PhysRevResearch.5.033007},
  url = {https://link.aps.org/doi/10.1103/PhysRevResearch.5.033007}
}

@article{wawrzik2023surface,
  title={{Surface induced electronic {Berry} curvature in bulk {Berry} curvature free materials}},
  author={Wawrzik, Dennis and Facio, Jorge I and van den Brink, Jeroen},
  journal={Mater. Today Phys.},
  volume={33},
  pages={101027},
  year={2023},
  url={https://www.sciencedirect.com/science/article/pii/S2542529323000639},
  publisher={Elsevier}
}

@article{Wawrzik2021,
    author = "Wawrzik, Dennis and You, Jhih-Shih and Facio, Jorge I. and van den Brink, Jeroen and Sodemann, Inti",
    title = "{Infinite Berry Curvature of Weyl Fermi Arcs}",
    volume = "127",
    ISSN = "1079-7114",
    url = "http://dx.doi.org/10.1103/PhysRevLett.127.056601",
    DOI = "10.1103/physrevlett.127.056601",
    number = "5",
    journal = "Phys. Rev. Lett.",
    publisher = "American Physical Society (APS)",
    year = "2021",
    month = "",
    pages = "056601"
}

@article{Chang1984,
    author = "Chang, Roger and Goddard, William A.",
    title = "{Reconstruction of the (110) surfaces for III–V semiconductors; Five systems involving In or Sb}",
    volume = "144",
    ISSN = "0039-6028",
    url = "http://dx.doi.org/10.1016/0039-6028(84)90103-1",
    DOI = "10.1016/0039-6028(84)90103-1",
    number = "2-3",
    journal = "Surf. Sci.",
    publisher = "Elsevier BV",
    year = "1984",
    month = "",
    pages = "311–320"
}

@article{Nienhaus1997,
    author = "Nienhaus, Hermann",
    title = "{Phonons in {110} surfaces of {III-V} compound semiconductors}",
    volume = "56",
    ISSN = "1095-3795",
    url = "http://dx.doi.org/10.1103/PhysRevB.56.13194",
    DOI = "10.1103/physrevb.56.13194",
    number = "20",
    journal = "Phys. Rev. B",
    publisher = "American Physical Society (APS)",
    year = "1997",
    month = "November",
    pages = "13194–13201"
}

@article{Duke1985,
    author = "Duke, C.B. and Paton, A.",
    title = "{The surface geometry of GaAs(110): A response}",
    volume = "164",
    ISSN = "0039-6028",
    url = "http://dx.doi.org/10.1016/0039-6028(85)90695-8",
    DOI = "10.1016/0039-6028(85)90695-8",
    number = "1",
    journal = "Surf. Sci.",
    publisher = "Elsevier BV",
    year = "1985",
    month = "December",
    pages = "L797–L806"
}

@article{Lubinsky1976,
    author = "Lubinsky, A. R. and Duke, C. B. and Lee, B. W. and Mark, P.",
    title = "{Semiconductor Surface Reconstruction: The Rippled Geometry of GaAs(110)}",
    volume = "36",
    ISSN = "0031-9007",
    url = "http://dx.doi.org/10.1103/PhysRevLett.36.1058",
    DOI = "10.1103/physrevlett.36.1058",
    number = "17",
    journal = "Phys. Rev. Lett.",
    publisher = "American Physical Society (APS)",
    year = "1976",
    month = "April",
    pages = "1058–1061"
}

@article{Masud1982,
    author = "Masud, N",
    title = "{Structure analysis of GaAs(110) surface at room temperature}",
    volume = "15",
    ISSN = "0022-3719",
    url = "http://dx.doi.org/10.1088/0022-3719/15/14/025",
    DOI = "10.1088/0022-3719/15/14/025",
    number = "14",
    journal = "J. Phys. C: Solid State Phys.",
    publisher = "IOP Publishing",
    year = "1982",
    month = "May",
    pages = "3209–3222"
}

@article{Yi2011,
    author = "Yi, Zhijun and Ma, Yuchen and Rohlfing, Michael",
    title = "{Silicon Donors at the GaAs(110) Surface: A First Principles Study}",
    volume = "115",
    ISSN = "1932-7455",
    url = "http://dx.doi.org/10.1021/jp2081403",
    DOI = "10.1021/jp2081403",
    number = "47",
    journal = "J. Phys. Chem. C",
    publisher = "American Chemical Society (ACS)",
    year = "2011",
    month = "November",
    pages = "23455–23462"
}

@article{Joannopoulos1974,
    author = "Joannopoulos, J. D. and Cohen, Marvin L.",
    title = "{Intrinsic surface states of (110) surfaces of group {IV} and {III-V} semiconductors}",
    volume = "10",
    ISSN = "0556-2805",
    url = "http://dx.doi.org/10.1103/PhysRevB.10.5075",
    DOI = "10.1103/physrevb.10.5075",
    number = "12",
    journal = "Phys. Rev. B",
    publisher = "American Physical Society (APS)",
    year = "1974",
    month = "December",
    pages = "5075–5081"
}

@article{Sabisch1995,
    author = "Sabisch, Magdalena and Krüger, Peter and Pollmann, Johannes",
    title = "{Ab initio calculations of SiC(110) and GaAs(110) surfaces: A comparative study and the role of ionicity}",
    volume = "51",
    ISSN = "1095-3795",
    url = "http://dx.doi.org/10.1103/PhysRevB.51.13367",
    DOI = "10.1103/physrevb.51.13367",
    number = "19",
    journal = "Phys. Rev. B",
    publisher = "American Physical Society (APS)",
    year = "1995",
    month = "May",
    pages = "13367–13380"
}

@article{Engels1998,
    author = "Engels, B. and Richard, P. and Schroeder, K. and Blügel, S. and Ebert, Ph. and Urban, K.",
    title = "{Comparison between ab initio theory and scanning tunneling microscopy for (110) surfaces of {III-V} semiconductors}",
    volume = "58",
    ISSN = "1095-3795",
    url = "http://dx.doi.org/10.1103/PhysRevB.58.7799",
    DOI = "10.1103/physrevb.58.7799",
    number = "12",
    journal = "Phys. Rev. B",
    publisher = "American Physical Society (APS)",
    year = "1998",
    month = "",
    pages = "7799–7815"
}

@article{Kulkova2005,
    author = "Kulkova, S.E. and Khanin, D.V. and Subashiev, A.V.",
    title = "{Simulation of Na, K, Cs and Cs(O) adsorption on GaAs(1 1 0) and (1 0 0) surfaces: towards predictable electronic structure of activation layer}",
    volume = "536",
    ISSN = "0168-9002",
    url = "http://dx.doi.org/10.1016/j.nima.2004.08.087",
    DOI = "10.1016/j.nima.2004.08.087",
    number = "3",
    journal = "Nucl. Instrum. Methods Phys. Res. A",
    publisher = "Elsevier BV",
    year = "2005",
    month = "January",
    pages = "295–301"
}

@article{Aroyo2006,
    author = "Aroyo, Mois Ilia and Perez-Mato, Juan Manuel and Capillas, Cesar and Kroumova, Eli and Ivantchev, Svetoslav and Madariaga, Gotzon and Kirov, Asen and Wondratschek, Hans",
    title = "{Bilbao Crystallographic Server: I. Databases and crystallographic computing programs}",
    volume = "221",
    ISSN = "2194-4946",
    url = "http://dx.doi.org/10.1524/zkri.2006.221.1.15",
    DOI = "10.1524/zkri.2006.221.1.15",
    number = "1",
    journal = "Z. Kristallogr. – Cryst. Mater.",
    publisher = "Walter de Gruyter GmbH",
    year = "2006",
    month = "January",
    pages = "15–27"
}

@article{Koepernik1999,
    author = "Koepernik, Klaus and Eschrig, Helmut",
    title = "{Full-potential nonorthogonal local-orbital minimum-basis band-structure scheme}",
    volume = "59",
    ISSN = "1095-3795",
    url = "http://dx.doi.org/10.1103/PhysRevB.59.1743",
    DOI = "10.1103/physrevb.59.1743",
    number = "3",
    journal = "Phys. Rev. B",
    publisher = "American Physical Society (APS)",
    year = "1999",
    month = "January",
    pages = "1743–1757"
}

@article{Opahle1999,
    author = "Opahle, I. and Koepernik, K. and Eschrig, H.",
    title = "{Full-potential band-structure calculation of iron pyrite}",
    volume = "60",
    ISSN = "1095-3795",
    url = "http://dx.doi.org/10.1103/PhysRevB.60.14035",
    DOI = "10.1103/physrevb.60.14035",
    number = "20",
    journal = "Phys. Rev. B",
    publisher = "American Physical Society (APS)",
    year = "1999",
    month = "November",
    pages = "14035–14041"
}

@article{Koepernik2023,
    author = "Koepernik, K. and Janson, O. and Sun, Yan and van den Brink, J.",
    title = "{Symmetry-conserving maximally projected Wannier functions}",
    volume = "107",
    ISSN = "2469-9969",
    url = "http://dx.doi.org/10.1103/PhysRevB.107.235135",
    DOI = "10.1103/physrevb.107.235135",
    number = "23",
    journal = "Phys. Rev. B",
    publisher = "American Physical Society (APS)",
    year = "2023",
    month = "",
    pages = "235135"
}

@article{Blakemore1982,
    author = "Blakemore, J. S.",
    title = "{Semiconducting and other major properties of gallium arsenide}",
    volume = "53",
    ISSN = "1089-7550",
    url = "http://dx.doi.org/10.1063/1.331665",
    DOI = "10.1063/1.331665",
    number = "10",
    journal = "J. Appl. Phys.",
    publisher = "AIP Publishing",
    year = "1982",
    month = "October",
    pages = "R123–R181"
}

@article{Perdew1983,
    author = "Perdew, John P. and Levy, Mel",
    title = "{Physical Content of the Exact Kohn-Sham Orbital Energies: Band Gaps and Derivative Discontinuities}",
    volume = "51",
    ISSN = "0031-9007",
    url = "http://dx.doi.org/10.1103/PhysRevLett.51.1884",
    DOI = "10.1103/physrevlett.51.1884",
    number = "20",
    journal = "Phys. Rev. Lett.",
    publisher = "American Physical Society (APS)",
    year = "1983",
    month = "November",
    pages = "1884–1887"
}

@article{Ma2016,
    author = "Ma, Jie and Wang, Lin-Wang",
    title = "{Using Wannier functions to improve solid band gap predictions in density functional theory}",
    volume = "6",
    ISSN = "2045-2322",
    url = "http://dx.doi.org/10.1038/srep24924",
    DOI = "10.1038/srep24924",
    number = "1",
    journal = "Sci. Rep.",
    publisher = "Springer Science and Business Media LLC",
    year = "2016",
    month = "April",
    pages = "24924"
}

@article{Jiang2020,
    author = "Jiang, M. and Xiao, H. Y. and Peng, S. M. and Qiao, L. and Yang, G. X. and Liu, Z. J. and Zu, X. T.",
    title = "{Effects of stacking periodicity on the electronic and optical properties of GaAs/AlAs superlattice: a first-principles study}",
    volume = "10",
    ISSN = "2045-2322",
    url = "http://dx.doi.org/10.1038/s41598-020-61509-x",
    DOI = "10.1038/s41598-020-61509-x",
    number = "1",
    journal = "Sci. Rep.",
    publisher = "Springer Science and Business Media LLC",
    year = "2020",
    month = "March",
    pages = "4862"
}

@article{Garcia-Moliner1976,
    author = "Garcia-Moliner, F and Flores, F",
    title = "{Theory of electronic surface states in semiconductors}",
    volume = "9",
    ISSN = "0022-3719",
    url = "http://dx.doi.org/10.1088/0022-3719/9/9/005",
    DOI = "10.1088/0022-3719/9/9/005",
    number = "9",
    journal = "J. Phys. C: Solid State Phys.",
    publisher = "IOP Publishing",
    year = "1976",
    month = "May",
    pages = "1609–1634"
}

@article{Kahn1978,
    author = "Kahn, A. and So, E. and Mark, P. and Duke, C. B. and Meyer, R. J.",
    title = "{Surface and near-surface atomic structure of GaAs (110)}",
    volume = "15",
    ISSN = "0022-5355",
    url = "http://dx.doi.org/10.1116/1.569697",
    DOI = "10.1116/1.569697",
    number = "4",
    journal = "J. Vac. Sci. Technol.",
    publisher = "American Vacuum Society",
    year = "1978",
    month = "",
    pages = "1223–1228"
}

@article{Qian1988,
    author = "Qian, Guo-Xin and Martin, Richard M. and Chadi, D. J.",
    title = "{First-principles calculations of atomic and electronic structure of the GaAs(110) surface}",
    volume = "37",
    ISSN = "0163-1829",
    url = "http://dx.doi.org/10.1103/PhysRevB.37.1303",
    DOI = "10.1103/physrevb.37.1303",
    number = "3",
    journal = "Phys. Rev. B",
    publisher = "American Physical Society (APS)",
    year = "1988",
    month = "January",
    pages = "1303–1307"
}

@article{Anua2012,
    author = {Anua, N. Najwa and Ahmed, R. and Saeed, M. A. and Shaari, A. and Haq, Bakhtiar Ul},
    title = {{DFT} investigations of structural and electronic properties of gallium arsenide ({GaAs})},
    journal = {AIP Conf. Proc.},
    volume = {1482},
    number = {1},
    pages = {64-68},
    year = {2012},
    month = {09},
    issn = {0094-243X},
    doi = {10.1063/1.4757439},
    url = {https://doi.org/10.1063/1.4757439}
}

@article{Ahmed2009,
author = {Ahmed, Rashid and Fazal-E-Aleem and Rashid, Haris and Akbarzadeh, H. and Hashemifar, S. Javad},

title = {STRUCTURAL PROPERTIES OF {III}-NITRIDE BINARY COMPOUNDS: A COMPREHENSIVE STUDY},
journal = {Mod. Phys. Lett. B},
volume = {23},
number = {08},
pages = {1111-1127},
year = {2009},
doi = {10.1142/S0217984909019247},

URL = { 
    
        https://doi.org/10.1142/S0217984909019247
    
    

}
}

@article{Wang2002,
doi = {10.1088/0953-8984/14/41/313},
url = {https://doi.org/10.1088/0953-8984/14/41/313},
year = {2002},
month = {oct},
publisher = {},
volume = {14},
number = {41},
pages = {9579},
author = {S Q Wang and H Q Ye},
title = {A plane-wave pseudopotential study on {III–V} zinc-blende and wurtzite
semiconductors under pressure},
journal = {J. Phys.: Condens. Matter}
}

@article{Rushton2001,
  title = {Density-functional calculations of semiconductor properties using a semiempirical exchange-correlation functional},
  author = {Rushton, Philip P. and Clark, Stewart J. and Tozer, David J.},
  journal = {Phys. Rev. B},
  volume = {63},
  issue = {11},
  pages = {115206},
  numpages = {5},
  year = {2001},
  month = {Mar},
  publisher = {American Physical Society},
  doi = {10.1103/PhysRevB.63.115206},
  url = {https://link.aps.org/doi/10.1103/PhysRevB.63.115206}
}

@article{Kalvoda1997,
  title = {Influence of electron correlations on ground-state properties of {III-V} semiconductors},
  author = {Kalvoda, Simon and Paulus, Beate and Fulde, Peter and Stoll, Hermann},
  journal = {Phys. Rev. B},
  volume = {55},
  issue = {7},
  pages = {4027--4030},
  numpages = {0},
  year = {1997},
  month = {Feb},
  publisher = {American Physical Society},
  doi = {10.1103/PhysRevB.55.4027},
  url = {https://link.aps.org/doi/10.1103/PhysRevB.55.4027}
}

@article{Causa1991,
  title = {Pseudopotential Hartree-Fock study of seventeen  and {IV-IV} semiconductors},
  author = {Caus\`a, M. and Dovesi, R. and Roetti, C.},
  journal = {Phys. Rev. B},
  volume = {43},
  issue = {14},
  pages = {11937--11943},
  numpages = {0},
  year = {1991},
  month = {May},
  publisher = {American Physical Society},
  doi = {10.1103/PhysRevB.43.11937},
  url = {https://link.aps.org/doi/10.1103/PhysRevB.43.11937}
}

@article{537j-v1lp,
  title = {Band geometry induced third-harmonic generation},
  author = {Sarkar, Sanjay and Mandal, Debottam and Agarwal, Amit},
  journal = {Phys. Rev. B},
  volume = {112},
  issue = {24},
  pages = {245425},
  numpages = {15},
  year = {2025},
  month = {Dec},
  publisher = {American Physical Society},
  doi = {10.1103/537j-v1lp},
  url = {https://link.aps.org/doi/10.1103/537j-v1lp}
}

@article{PhysRevLett.129.227401,
  title = {Resonant Second-Harmonic Generation as a Probe of Quantum Geometry},
  author = {Bhalla, Pankaj and Das, Kamal and Culcer, Dimitrie and Agarwal, Amit},
  journal = {Phys. Rev. Lett.},
  volume = {129},
  issue = {22},
  pages = {227401},
  numpages = {8},
  year = {2022},
  month = {Nov},
  publisher = {American Physical Society},
  doi = {10.1103/PhysRevLett.129.227401},
  url = {https://link.aps.org/doi/10.1103/PhysRevLett.129.227401}
}

@article{nagaosa2017,
author = {Nagaosa, Naoto and Morimoto, Takahiro},
title = {Concept of Quantum Geometry in Optoelectronic Processes in Solids: Application to Solar Cells},
journal = {Adv. Mater.},
volume = {29},
number = {25},
pages = {1603345},
doi = {https://doi.org/10.1002/adma.201603345},
url = {https://advanced.onlinelibrary.wiley.com/doi/abs/10.1002/adma.201603345},
year = {2017}
}

@article{sipe2000,
  title = {Second-order optical response in semiconductors},
  author = {Sipe, J. E. and Shkrebtii, A. I.},
  journal = {Phys. Rev. B},
  volume = {61},
  issue = {8},
  pages = {5337--5352},
  numpages = {0},
  year = {2000},
  month = {Feb},
  publisher = {American Physical Society},
  doi = {10.1103/PhysRevB.61.5337},
  url = {https://link.aps.org/doi/10.1103/PhysRevB.61.5337}
}

@misc{yang2017,
  author  = {X. Yang and K. Burch and Y. Ran},
  title   = {Divergent Bulk Photovoltaic Effect in {Weyl} Semimetals},
  year    = {2017},
  eprint  = {1712.09363},
  archivePrefix = {arXiv},
  primaryClass = {cond-mat.mtrl-sci},
  note    = {arXiv preprint}
}

@article{PhysRevX.10.041041,
  title = {Low-Frequency Divergence and Quantum Geometry of the Bulk Photovoltaic Effect in Topological Semimetals},
  author = {Ahn, Junyeong and Guo, Guang-Yu and Nagaosa, Naoto},
  journal = {Phys. Rev. X},
  volume = {10},
  issue = {4},
  pages = {041041},
  numpages = {28},
  year = {2020},
  month = {Nov},
  publisher = {American Physical Society},
  doi = {10.1103/PhysRevX.10.041041},
  url = {https://link.aps.org/doi/10.1103/PhysRevX.10.041041}
}

@article{PhysRevLett.121.246403,
  title = {{Strongly Enhanced {Berry} Dipole at Topological Phase Transitions in BiTeI}},
  author = {Facio, Jorge I. and Efremov, Dmitri and Koepernik, Klaus and You, Jhih-Shih and Sodemann, Inti and van den Brink, Jeroen},
  journal = {Phys. Rev. Lett.},
  volume = {121},
  issue = {24},
  pages = {246403},
  numpages = {6},
  year = {2018},
  month = {Dec},
  publisher = {American Physical Society},
  doi = {10.1103/PhysRevLett.121.246403},
  url = {https://link.aps.org/doi/10.1103/PhysRevLett.121.246403}
}

@article{https://doi.org/10.1002/qute.202100056,
author = {Ortix, Carmine},
title = {{Nonlinear Hall Effect with Time-Reversal Symmetry: Theory and Material Realizations}},
journal = {Adv. Quantum Technol.},
volume = {4},
number = {9},
pages = {2100056},
url = {https://advanced.onlinelibrary.wiley.com/doi/abs/10.1002/qute.202100056},
year = {2021}
}

@ARTICLE{Kang2019-ek,
  title     = "Nonlinear anomalous Hall effect in few-layer {WTe$_2$}",
  author    = "Kang, Kaifei and Li, Tingxin and Sohn, Egon and Shan, Jie and
               Mak, Kin Fai",
  journal   = "Nat. Mater.",
  publisher = "Springer Science and Business Media LLC",
  volume    =  18,
  number    =  4,
  pages     = "324--328",
  month     =  apr,
  year      =  2019,
  url = "https://www.nature.com/articles/s41563-019-0294-7"
}

@article{PhysRevLett.134.066603,
  title = {{Observation of Temperature-Independent Anomalous Hall Effect in Thin Bismuth from Near Absolute Zero to 300 K Temperature}},
  author = {Yu, Oulin and Boivin, F. and Silberztein, A. and Gervais, G.},
  journal = {Phys. Rev. Lett.},
  volume = {134},
  issue = {6},
  pages = {066603},
  numpages = {6},
  year = {2025},
  month = {Feb},
  publisher = {American Physical Society},
  doi = {10.1103/PhysRevLett.134.066603},
  url = {https://link.aps.org/doi/10.1103/PhysRevLett.134.066603}
}

@article{PhysRevB.97.041101,
  title = {{Berry curvature dipole in {Weyl} semimetal materials: An ab initio study}},
  author = {Zhang, Yang and Sun, Yan and Yan, Binghai},
  journal = {Phys. Rev. B},
  volume = {97},
  issue = {4},
  pages = {041101(R)},
  numpages = {6},
  year = {2018},
  month = {Jan},
  publisher = {American Physical Society},
  doi = {10.1103/PhysRevB.97.041101},
  url = {https://link.aps.org/doi/10.1103/PhysRevB.97.041101}
}

@article{PhysRevLett.121.266601,
  title = {{Band Signatures for Strong Nonlinear Hall Effect in Bilayer ${\mathrm{WTe}}_{2}$}},
  author = {Du, Z. Z. and Wang, C. M. and Lu, Hai-Zhou and Xie, X. C.},
  journal = {Phys. Rev. Lett.},
  volume = {121},
  issue = {26},
  pages = {266601},
  numpages = {6},
  year = {2018},
  month = {Dec},
  publisher = {American Physical Society},
  doi = {10.1103/PhysRevLett.121.266601},
  url = {https://link.aps.org/doi/10.1103/PhysRevLett.121.266601}
}

@article{PhysRevB.98.121109,
  title = {{Berry} curvature dipole current in the transition metal dichalcogenides family},
  author = {You, Jhih-Shih and Fang, Shiang and Xu, Su-Yang and Kaxiras, Efthimios and Low, Tony},
  journal = {Phys. Rev. B},
  volume = {98},
  issue = {12},
  pages = {121109(R)},
  numpages = {6},
  year = {2018},
  month = {Sep},
  publisher = {American Physical Society},
  doi = {10.1103/PhysRevB.98.121109},
  url = {https://link.aps.org/doi/10.1103/PhysRevB.98.121109}
}

@ARTICLE{Ma2019-qi,
  title     = "Observation of the nonlinear Hall effect under
               time-reversal-symmetric conditions",
  author    = "Ma, Qiong and Xu, Su-Yang and Shen, Huitao and MacNeill, David
               and Fatemi, Valla and Chang, Tay-Rong and Mier Valdivia,
               Andr{\'e}s M and Wu, Sanfeng and Du, Zongzheng and Hsu,
               Chuang-Han and Fang, Shiang and Gibson, Quinn D and Watanabe,
               Kenji and Taniguchi, Takashi and Cava, Robert J and Kaxiras,
               Efthimios and Lu, Hai-Zhou and Lin, Hsin and Fu, Liang and
               Gedik, Nuh and Jarillo-Herrero, Pablo",
  
  journal   = "Nature",
  publisher = "Springer Science and Business Media LLC",
  volume    =  565,
  number    =  7739,
  pages     = "337--342",
  month     =  jan,
  year      =  2019,
  url = "https://www.nature.com/articles/s41586-018-0807-6"
}

@article{Tong1978,
  title = {Surface bond angle and bond lengths of rearranged As and Ga atoms on {GaAs}(110)},
  author = {Tong, S. Y. and Lubinsky, A. R. and Mrstik, B. J. and Van Hove, M. A.},
  journal = {Phys. Rev. B},
  volume = {17},
  issue = {8},
  pages = {3303--3309},
  numpages = {0},
  year = {1978},
  month = {Apr},
  publisher = {American Physical Society},
  doi = {10.1103/PhysRevB.17.3303},
  url = {https://link.aps.org/doi/10.1103/PhysRevB.17.3303}
}

@article{Smit1985,
title = {The relaxed {GaAs}(110) surface: Are bond-lengths conserved?},
journal = {Surf. Sci.},
volume = {150},
number = {1},
pages = {245-251},
year = {1985},
issn = {0039-6028},
doi = {https://doi.org/10.1016/0039-6028(85)90221-3},
url = {https://www.sciencedirect.com/science/article/pii/0039602885902213},
author = {L. Smit and T.E. Derry and J.F. {Van Der Veen}}
}

@article{duke1983,
title = {The atomic geometry of {GaAs}(110) revisited},
journal = {Surf. Sci.},
volume = {127},
number = {2},
pages = {L135-L143},
year = {1983},
issn = {0039-6028},
doi = {https://doi.org/10.1016/0039-6028(83)90412-0},
url = {https://www.sciencedirect.com/science/article/pii/0039602883904120},
author = {C.B. Duke and S.L. Richardson and A. Paton and A. Kahn}
}

@article{PhysRevLett.79.1917,
  title = {Atomically Precise {GaAs/AlGaAs} Quantum Dots Fabricated by Twofold Cleaved Edge Overgrowth},
  author = {Wegscheider, W. and Schedelbeck, G. and Abstreiter, G. and Rother, M. and Bichler, M.},
  journal = {Phys. Rev. Lett.},
  volume = {79},
  issue = {10},
  pages = {1917--1920},
  numpages = {0},
  year = {1997},
  month = {Sep},
  publisher = {American Physical Society},
  doi = {10.1103/PhysRevLett.79.1917},
  url = {https://link.aps.org/doi/10.1103/PhysRevLett.79.1917}
}

@article{Rössler_2010,
doi = {10.1088/1367-2630/12/4/043007},
url = {https://doi.org/10.1088/1367-2630/12/4/043007},
year = {2010},
month = {apr},
publisher = {},
volume = {12},
number = {4},
pages = {043007},
author = {Rössler, C and Feil, T and Mensch, P and Ihn, T and Ensslin, K and Schuh, D and Wegscheider, W},
title = {Gating of high-mobility two-dimensional electron gases in {GaAs/AlGaAs} heterostructures},
journal = {New J. Phys.}
}

@article{SCHLAPFER2016114,
title = {Photoluminescence and the gallium problem for highest-mobility {GaAs/AlGaAs}-based {2D} electron gases},
journal = {J. Cryst. Growth},
volume = {442},
pages = {114-120},
year = {2016},
issn = {0022-0248},
doi = {https://doi.org/10.1016/j.jcrysgro.2016.02.039},
url = {https://www.sciencedirect.com/science/article/pii/S0022024816300720},
author = {F. Schläpfer and W. Dietsche and C. Reichl and S. Faelt and W. Wegscheider}
}

@article{PhysRevLett.117.256601,
  title = {Linear Magnetoresistance in a Quasifree Two-Dimensional Electron Gas in an Ultrahigh Mobility {GaAs} Quantum Well},
  author = {Khouri, T. and Zeitler, U. and Reichl, C. and Wegscheider, W. and Hussey, N. E. and Wiedmann, S. and Maan, J. C.},
  journal = {Phys. Rev. Lett.},
  volume = {117},
  issue = {25},
  pages = {256601},
  numpages = {4},
  year = {2016},
  month = {Dec},
  publisher = {American Physical Society},
  doi = {10.1103/PhysRevLett.117.256601},
  url = {https://link.aps.org/doi/10.1103/PhysRevLett.117.256601}
}

@article{
doi:10.1126/science.1216022,
author = {G. Scalari  and C. Maissen  and D. Turčinková  and D. Hagenmüller  and S. De Liberato  and C. Ciuti  and C. Reichl  and D. Schuh  and W. Wegscheider  and M. Beck  and J. Faist },
title = {Ultrastrong Coupling of the Cyclotron Transition of a {2D} Electron Gas to a {THz} Metamaterial},
journal = {Science},
volume = {335},
number = {6074},
pages = {1323-1326},
year = {2012},
doi = {10.1126/science.1216022},
URL = {https://www.science.org/doi/abs/10.1126/science.1216022}}

@article{Külah_2021,
doi = {10.1088/1361-6641/ac0d98},
url = {https://doi.org/10.1088/1361-6641/ac0d98},
year = {2021},
month = {jul},
publisher = {IOP Publishing},
volume = {36},
number = {8},
pages = {085013},
author = {Külah, E and Reichl, C and Scharnetzky, J and Alt, L and Dietsche, W and Wegscheider, W},
title = {The improved inverted {AlGaAs/GaAs} interface: its relevance for high-mobility quantum wells and hybrid systems},
journal = {Semicond. Sci. Technol.}
}

@misc{Blatter2026,
  author    = {Blatter, G. and van den Brink, J. and Facio, J.I},
  title     = {Dataset for: Surface-bulk hybridization enhances the Berry curvature dipole in {GaAs}(110)},
  publisher = {Zenodo},
  year      = {2026},
  doi       = {10.5281/zenodo.22659646},
  url       = {https://doi.org/10.5281/zenodo.22659646}
}

\appendix
\setcounter{secnumdepth}{2}

\renewcommand{\thesection}{\Alph{section}}      
\renewcommand{\thesubsection}{\thesection.\arabic{subsection}}  

\end{document}